\documentstyle[12pt]{article}
\renewcommand\theequation{\thesection.\arabic{equation}}
\begin{document}
\setlength{\unitlength}{1mm}
\begin{center}
{\Large\bf Nonabelian Gauge Antisymmetric Tensor Fields}
\end{center}
\begin{center}
{\large\bf S.N.Solodukhin}
\end{center}
\begin{center}
{\bf Department of Theoretical Physics, Physics Faculty of Moscow
University, Moscow 117234, Russia}
\end{center}
\vspace*{2cm}
\abstract

   We construct the theory of non-abelian gauge antisymmetric tensor
fields, which generalize the standard Yang-MIlls fields and abelian
gauge p-forms. The corresponding gauge group acts on the space of
inhomogeneous differential forms and it is shown to be a supergroup.

   We construct the theory of non-abelian gauge antisymmetric tensor
fields, which generalize the standard Yang-MIlls fields and abelian
gauge p-forms. The corresponding gauge group acts on the space of
inhomogeneous differential forms and it is shown to be a supergroup.
The wide class of generalized Chern-Simons actions is constructed.
The wide class of generalized Chern-Simons actions is constructed.
\vspace*{2cm}
\section{Introduction}
\setcounter{equation}0
   Recently much attention has been payed to the study of theoretical
models with abelian gauge antisymmetric tensor fields (AGATF). It
turned out that AGATF play an important role in the string theory,
supergravity and gauge theory of gravity [1]. An important
circumstance is that the AGATF can be described using differential-
geometric methods as differential forms (in which case the parameters
of gauge transformations are also differential forms). On the other
hand, the fermionic matter fields can be described with help of a set
of antisymmetric tensor fields of arbitrary rank (i.e. inhomogeneous
differential forms) [2,3].

   During the last years the surprising relation of gauge fields to
the space-time topology was discovered. In the works of Donaldson [4]
and Witten [5] such a relation was demonstrated for the standard gauge
fields. On the other hand it was shown by Horowitz [6], and earlier by
Schwarz [7], that AGATF give us an example of so-called "topological
field theory" and lead to appearance of topological invariants (such
as the Ray-Singer torsion and linking numbers) under quantization. In
[6] the minimal non-abelian generalization of AGATF was also
considered. However, so far there doesn't exist a satisfactory
generalization of AGATF for non-abelian case. Attempts of such a
generalization [8] result in an essentially nonlocal and nonlinear
theory. Hence an actual problem is to construct a theory of
non-abelian gauge antisymmetric tensor fields (NGATF). These fields
could find numerous applications in many theoretical models where so
far one has used the AGATF. There is also a hope that NGATF will give
us new tools for the study of space-time topology. It is a goal of
this paper to attempt in construction of NGATF theory.

   At present we know two kinds of gauge fields. First is the standard
(Yang-Mills) non-abelian gauge field which is described as the Lie
algebra valued one-form  $A_{(1)}$  with the infinitesimal gauge
transformation law:
\begin{equation}
\delta_{\alpha} A_{(1)} = - d\alpha_{(0)} - A_{(1)} \alpha_{(0)} + \alpha_{(0)}
A_{(1)}
\label{eq:1.1}
\end{equation}
where $A_{(1)}=A^{a}_{\mu} \lambda^{a}dx^{\mu}$ , matrices $\lambda^{a}$
are the generators of a gauge group G, $\alpha_{(0)}=\alpha^{a} \lambda^{a}$
is an infinitesimal transformation parameter - the zero-form with values
in the Lie algebra of the group G.

   Second is the theory of AGATF which are n-forms  $A_{(n)}$  with the
gauge transformation law:
\begin{equation}
\delta_{\alpha} A_{(n)} = - d\alpha_{(n-1)}
\label{eq:1.2}
\end{equation}
where  (n-1)-form  $\alpha_{(n-1)}$ is an infinitesimal transformation
parameter.
   Our aim is to construct a theory of NGATF which would be n-forms
$A_{(n)}$  with values in the Lie algebra of a group G:
\begin{equation}
A_{(n)}={1\over n!} A^{a}_{\mu_{1}...\mu_{n}} \lambda^{a} dx^{\mu_{1}}\wedge...
\wedge dx^{\mu_{n}}
\label{eq:1.3}
\end{equation}
Such a theory should generalize both the Yang-Mills model and AGATF
theory in that the transformation laws (1.1),(1.2) are recovered as
the particular cases of a gauge transformation law for fields (1.3).

   One could naively expect the following transformation law
for $A_{(n)}$ :
\begin{equation}
\delta_{\alpha} A_{(n)} \sim -d\alpha_{(n-1)} - A\wedge \alpha_{(n-1)}
+ \alpha_{(n-1)} \wedge A
\label{eq:1.4}
\end{equation}
where
\begin{equation}
\alpha_{(n-1)}= {1 \over (n-1)!} \alpha^{a}_{\mu_{1}...\mu_{n}} \lambda^{a}
dx^{\mu_{1}} \wedge...\wedge dx^{\mu_{n-1}}
\end{equation}
is an infinitesimal parameter (n-1)-form with values in the algebra of
group $G$.

   However the formula (1.4) is evidently senseless, since the last
terms in (1.4) are differential form of rank (2n-1), while the
left-hand side of (1.4) is the n-form. Only for the case n=1 this
formula has sense. In order to make sense of (1.4) let us consider a
set of gauge forms $A_{(n)}, A_{(1)}$ such that (1.4) is replaced by
\begin{equation}
\delta_{\alpha} A_{(n)}  \sim  -d\alpha_{(n-1)} - A_{(1)} \wedge \alpha_{(n-1)}
+ \alpha_{(n-1)} \wedge A_{(1)}
\label{eq:1.4'}
\end{equation}
Natural generalization appears to take a complete set of forms
$A_{(n)}, A_{(n-1)}, ..., A_{(p)}, ..., A_{(1)}$ with a transformation law
\begin{equation}
\delta_{\alpha} A_{(n)} \sim  -d\alpha_{(n-1)} - \sum_{k} A_{(k)} \wedge
\alpha_{(n-k)} + \alpha_{(n-k)} \wedge A_{(k)}
\label{eq:1.4'}
\end{equation}
where $\alpha_{(0)}, \alpha_{(1)}, ..., \alpha_{(p)}, ..., \alpha_{(n-1)}$ are
infinitesimal gauge parameters - the set of k-forms with values in algebra of
group G.

   In the standard Yang-Mills theory the gauge transformations act on
a multiplet of matter fields:
\begin{equation}
\delta_{\alpha}\phi^{i} = \alpha^{i}_{\ j} \phi^{j}
\label{eq:1.5}
\end{equation}
Both $\phi$ and parameters are zero-forms.

   A natural generalization is to describe the matter fields by
n-forms, and parameters by n-forms. However,
\begin{equation}
\phi^{i}_{(n)} \rightarrow \phi^{i}_{(n)} + \alpha^{i}_{(k)j} \wedge
\phi^{j}_{(n)}
\label{eq:1.6}
\end{equation}
transfers n-forms into (n+k)-forms. Hence one should consider the set
of forms of all possible ranks on a given manifold, i.e. the multiplet
of inhomogeneous differential forms:
\begin{equation}
\Psi^{i}= \sum_{k} {1 \over k!} \phi^{i}_{\mu_{1}...\mu_{k}} dx^{\mu_{1}}
\wedge ... \wedge dx^{\mu_{k}}
\label{eq:1.7}
\end{equation}
Then the transformation generated by inhomogeneous forms
\begin{equation}
\alpha^{i}_{\ j} = \sum_{k} {1 \over k!} \alpha^{i}_{\ j\mu_{1}...\mu_{k}}
dx^{\mu_{1}} \wedge... \wedge dx^{\mu_{k}}
\label{eq:1.8}
\end{equation}
acts on the field (1.10) as follows
\begin{equation}
\delta_{\alpha} \Psi^{i} = \alpha^{i}_{\ j} \wedge \Psi^{j}
\label{eq:1.9}
\end{equation}
Notice, that such a generalization was initiated by the study of
fermionic matter description in terms of inhomogeneous differential
forms [2,3]. In particular in [3] the example of supersymmetry
transformation mixing the forms of different rank was considered.

\section{Infinitesimal gauge transformations}
\setcounter{equation}0
   Let us consider an action for matter fields described by
differential form $\Psi$ taking values in an linear space ${\cal L}^{N}$
(1.10) which is invariant under the global transformation (1.12):
\begin{equation}
\Psi \rightarrow \Psi + \alpha \wedge \Psi
\label{eq:2.1}
\end{equation}
where the infinitesimal transformation parameter $\alpha$ (which is
matrix-valued inhomogeneous form (1.11) ) has constant coefficients.
The kinetic term of an action depends on $d\Psi$.

   Under global transformations (2.1) one has
\begin{equation}
d\Psi \rightarrow d\Psi + \eta \alpha \wedge d \Psi
\label{eq:2.2}
\end{equation}
where the operator $\eta$ acts on the forms as follows:
\begin{equation}
\eta \alpha = \sum_{k} {(-1)^{k} \over k!} \alpha_{\mu_{1}...\mu_{k}}
dx^{\mu_{1}} \wedge...\wedge dx^{\mu_{k}}
\label{eq:2.3}
\end{equation}
This is the involution, one easily sees that  $\eta^{2}=id$.
   Then under local transformations (2.1) (when coefficients
$\alpha_{\mu_{1}...\mu_{k}}$  depend on $x$) one obtains
\begin{equation}
d\Psi \rightarrow d\Psi + \eta \alpha \wedge d\Psi + d\alpha \wedge \Psi
\label{eq:2.4}
\end{equation}
In order to compensate the last term one should introduce the
gauge field - an matrix-valued inhomogeneous form $A$ (1.3):
\begin{equation}
A = \sum_{k} A_{(k)}
\end{equation}
and consider the covariant derivative:
\begin{equation}
\nabla \Psi = d \Psi + A \wedge \Psi
\end{equation}
which transforms under local transformations (2.1) as follows
\begin{equation}
\nabla \Psi \rightarrow \nabla \Psi + \eta \alpha \wedge \nabla \Psi
\end{equation}
Then the gauge invariant  action is obtained substituting external
derivative $d\Psi$  by covariant derivative  $\nabla \Psi$.

   From (2.7) one obtains the transformation law for the gauge
field $A$:
\begin{equation}
A \rightarrow A - d\alpha + \eta \alpha \wedge A - A\wedge \alpha
\end{equation}
In terms of the inhomogeneous components (2.5) this formula reads
\begin{equation}
A_{(n)} \rightarrow A_{(n)} - d\alpha_{(n-1)} + \sum_{k=0}^{n-1}
[(-1)^{k} \alpha_{(k)} \wedge A_{(n-k)} - A_{(n-k)} \wedge \alpha_{(k)}]
\end{equation}
This formula makes more precise our  preliminary discussion in the
Introduction (see (1.7)).

  It is easy to see that indeed the transformation law (2.8), (2.9)
contains the laws (1.1), (1.2) as particular cases.

   The curvature form for AGATF
\begin{equation}
F_{(n+1)} = dA_{(n)}
\end{equation}
is invariant under gauge transformation (1.2), while the curvature for
the standard Yang-Mills gauge field
\begin{equation}
F_{(2)} = dA_{(1)} + A_{(1)} \wedge A_{(1)}
\end{equation}
under a gauge transformations (1.1)  changes as
\begin{equation}
\delta_{\alpha} F_{(2)}= \alpha_{(0)} F_{(2)} - F_{(2)} \alpha_{(0)}
\end{equation}
   Let us find the generalized expression for the NGATF curvature with
the transformation law
\begin{equation}
F \rightarrow F + \alpha \wedge F - F \wedge \alpha
\end{equation}
It is easy to see that the proper generalization reads
\begin{equation}
F = dA - (\eta A) \wedge A
\end{equation}
This satisfies  all necessary conditions.

   For homogeneous components $F_{(k)}$ : $F= \sum^{ }_{k} F_{(k)}$ one has
\begin{equation}
F_{(n+1)} = dA_{(n)} - \sum_{k=1}^{n} (-1)^{k} A_{(k)} \wedge A_{(n-k+1)}
\end{equation}
It is also easy to verify the Jacobi identity for the curvature $F$ ,
\begin{equation}
dF + A \wedge F - (\eta F) \wedge A = 0
\end{equation}
   In the standard Yang-Mills theory one usually consider the gauge
one-form   $A_{(1)}$  to be antihermitean. However, there are no reasons
for such a restrictions in our general construction.

\section{ The gauge group}
\setcounter{equation}0
   Let us discuss now the group structure of transformations (1.12)
acting on the space of inhomogeneous forms $\Psi$ with values in a
N-dimensional linear space (see (1.10)).

   Let $G$ be an inhomogeneous form with values in a space of matrices
$GL(N,C)$:
\begin{eqnarray}
&&G= \sum_{k} G_{(k)}\nonumber\\
&&G_{(k)}= {1 \over k!} G^{i}_{\ j\mu_{1}...\mu_{k}} dx^{\mu_{1}}\wedge ...
\wedge dx^{\mu_{k}}
\end{eqnarray}
We determine the group multiplication of such forms as follows:
\begin{equation}
G^{\prime} \circ G^{\prime \prime} = G^{\prime i}_{\ j} \wedge
G^{\prime \prime j}_{\ \ k}
\end{equation}
In general case the exterior algebra is not a group, since not for
every element one can determine the inverse element with respect to
the wedge product $\wedge$. The existence condition for such an inverse
element is the nondegenerate scalar part $G_{(0)}\neq 0$ in abelian case
(when $\{G\}=\Lambda^{\ast} M$), and
\begin{equation}
det[G^{i}_{(0)j}]\neq 0
\end{equation}
in the non-abelian case.
Consequently one must consider the space of all forms (3.1) satisfying
the condition (3.3).

   In order to construct an action for the matter fields (1.10) it is
necessary to have an invariant scalar product on the space of
inhomogeneous differential forms (1.10). The standard bilinear form
\begin{equation}
\int_{ }^{ } \ast \Psi^{+} \wedge \Psi
\end{equation}

is invariant under transformations
\begin{equation}
\Psi \rightarrow G \wedge \Psi
\end{equation}

if $G$ has only a zero rank component. Notice that (3.4) demands a
metric on manifold M which enters via the Hodge operator $\ast$.

   Instead, let us consider the following scalar product for forms on
a D-dimensional manifold $M^{D}$ .
\begin{equation}
(\phi,\psi)= \int_{M^{D}} \xi \phi^{+} \wedge \psi = \sum_{k=0}^{D}
\int_{M^{D}} \xi \phi^{+}_{(k)} \wedge \psi_{(D-k)}
\end{equation}

where the operator $\xi$  acts on differential forms as follows
\begin{equation}
\xi \phi = \sum_{k=0}^{D} {1 \over k!} (-1)^{k(k-1) \over 2} \phi_{\mu_{1}
...\mu_{k}} dx^{\mu_{1}} \wedge ...dx^{\mu_{k}}
\end{equation}
Notice that $\xi$ is the anti-involution on  $\Lambda^{\ast}M^{D}$ ,
\begin{equation}
\xi (\phi \wedge \psi) = (\xi \psi) \wedge (\xi\phi)
\end{equation}

For invariance of (3.6) under transformations (3.5) it is sufficient
to set the condition
\begin{equation}
\xi G^{+} \wedge G =1
\end{equation}

Clearly, all such $G$ form a group.

   Let us consider
\begin{equation}
G= 1+\alpha
\end{equation}
where $\alpha$ is an infinitesimal differential form.

   Then from (3.9) one finds
\begin{equation}
\xi \alpha^{+} = - \alpha
\end{equation}
General solution of this equation reads
\begin{equation}
\alpha = \sum_{k=0}^{D}{ (\imath)^{k(k-1) \over 2} \over k!} ( \rho^{a}_{(k)}
\lambda^{a} + \rho_{(k)} \imath I )
\end{equation}
where $\rho^{ \bar{a}}_{(k)}= (\rho_{(k)}, \rho^{a}_{(k)})$ are k-forms with
real coefficients, and $N\times N$ matrices $\lambda^{\bar{a}}=(\imath I,
\lambda^{a})$  are the antihermitean generators of the unitary group $U(N)$
with the structure constants $f^{\bar{a} \bar{b} \bar{c}}$ . It should be
noted that there is no additional condition (like $\det g =1$ for usual
unitary group) for group elements $G$ (3.9) in order to take away abelian
subgroup generated by $\rho_{(k)} \imath I$ (3.12).

   Thus the generators of the group (3.10) has the form
\begin{eqnarray}
&&X^{\bar{a}}_{(s)} = \lambda^{\bar{a}} I_{(s)},\nonumber\\
&&I_{(s)} = {1 \over k!} dx^{\mu_{1}}\wedge ... \wedge dx^{\mu_{k}}, k=0,...,D
\end{eqnarray}
These generators satisfy
\begin{equation}
X^{\bar{a}}_{(s)} X^{\bar{b}}_{(s^{\prime})} - (-1)^{|s||s^{\prime}|}
X^{\bar{b}}_{(s^{\prime})}
X^{\bar{a}}_{(s)} = f^{\bar{a} \bar{b} \bar{c}} X^{\bar{c}}_{(s+s^{\prime})}
\end{equation}
and so they define a superalgebra. Hence the group (3.10) is a
supergroup. Roughly, it is unitary group $U(N)$ with inhomogeneous
differential forms as parameters. Notice, that the supergroups appear
in ref.[9] in a similar manner.

   In Appendix we find the general form for a group $G$ elements
satisfying equation (3.9):
\begin{equation}
G=g_{0} (1 + H + {\cal F} )
\end{equation}
where zero-form $g_{0}$ is an element of the unitary group $U(N)$ :
\begin{equation}
g^{+}_{0} g_{0}=1
\end{equation}
and $H$ and $\cal F$   are inhomogeneous forms of rank $\geq 1$ which
satisfy the conditions:
\begin{eqnarray}
&&\xi H^{+} = - H \nonumber\\
&&\xi {\cal F}^{+} = {\cal F}
\end{eqnarray}
The form $\cal F$   is constructed as a polynomial of $H$ (see Appendix):
${\cal F} = \sum {a_{k} H^{2k}}$ relatively of wedge product.

   We can also define the subgroup of group $G$ (3.15) as follows:
\begin{equation}
G^{D}_{K} \equiv \{ G=g_{0} (1 + H + {\cal F}) | H_{(2k-1)} = 0,
k= 1, .., K  \}
\end{equation}
This subgroup will be useful in the next Section to write the
Chern-Simons terms for NGATF.

   At the end of this Section let us write the explicit matter field
actions invariant under global transformations (3.5), (3.9):
\begin{eqnarray}
&&S= \int_{M^{D}} \eta \tilde{\psi} \wedge d\phi \nonumber\\
&&S= \int_{M^{D}}d\tilde{\psi} \wedge d\phi
\end{eqnarray}
where
\begin{equation}
\tilde{\psi} = \xi {\psi}^{+}
\end{equation}
The gauge invariant version of these actions is
\begin{eqnarray}
&&S= \int_{M^{D}} \eta \tilde{\psi} \wedge \nabla \phi \nonumber\\
&&S= \int_{M^{D}}\nabla \tilde{\psi} \wedge \nabla \phi
\end{eqnarray}
where   $\nabla$  is the covariant derivative (2.6).

  Notice that we do not use metric on the space-time manifold $M^{D}$ to
write the actions (3.19),(3.21). Hence one can expect to obtain the
description of some topological invariants as quantum observables in
the models (3.19), (3.21).

\section{ Generalized Chern-Simons for NGATF}
\setcounter{equation}0

   In order to treat NGATF as a dynamical field one must formulate the
action for its description. It should be noted that the usual action
\begin{equation}
\int Tr( \ast F \wedge F)
\end{equation}
which is invariant in the abelian case cannot be directly generalized
to the non-abelian case. This was the main problem in earlier works on
NGATF (see [8]). In our formalism there is a natural possibility to
overcome the difficulty with the help of above introduced invariant
scalar product for inhomogeneous differential forms (3.6). One can
suggest the two expressions for the NGATF action:
\begin{equation}
( F, F ) = \int_{M^{D}} Tr ( \tilde{F} \wedge F)
\end{equation}
and
\begin{equation}
(\tilde{F}, F ) = \int_{M^{D}} Tr (F\wedge F)
\end{equation}
It is easy to see that both these expressions are invariant with
respect to the gauge transformations (2.13) if the space-time
dimension D is odd. However if D is even we have for the variation of
(4.2), (4.3):
\begin{equation}
\Delta_{\alpha} (F, F) = \int Tr [ (\alpha - \eta \alpha) \wedge F
\wedge F]
\end{equation}
\begin{equation}
\Delta_{\alpha} (\tilde{F}, F) = \int Tr [( \alpha - \eta \alpha) \wedge
\tilde{F} \wedge F]
\end{equation}
Consequently, in order to (4.2), (4.3) be gauge invariant, the
infinitesimal form $\alpha$  should be even:
\begin{equation}
\eta \alpha = \alpha
\end{equation}
More precisely, since the curvature form $F$ has rank $\geq 2$ , so the
homogeneous components of $\alpha$ should satisfy condition:
\begin{equation}
\alpha_{(2k+1)} = 0, k= 0, ..., {D \over 2} - 3
\end{equation}
In other words for even space-time dimension D the gauge group for
(4.2-3) is $G^{D}_{{D \over 2} - 3}$  (see (3.18)).

  Here we can restrict ourselves to the particular case in even D (see
also [10]):
\begin{eqnarray}
&&\eta A = - A,\nonumber\\
&&\eta F = F
\end{eqnarray}
Then from action (4.2) one obtains the equations of motion:
\begin{eqnarray}
&&d \tilde{F} + A \wedge \tilde{F} -(\eta \tilde{F}) \wedge A = 0 ,
{\rm (for\ odd\ D)}\nonumber\\
&&d \tilde{F} + A \wedge \tilde{F} - \tilde{F} \wedge A = 0 ,
{\rm (for\ even\ D)}
\end{eqnarray}
It is easy to see that if the curvature F satisfy to the one of the
conditions
\begin{eqnarray}
&&\tilde{F} = + F\nonumber\\
&&\tilde{F} = - F
\end{eqnarray}
then the equations (4.9) hold identically as a consequence of the
Jacobi identity (2.16). These conditions (4.10) are similar to the
self- and antiself-duality conditions for instantons in the standard
Yang-Mills theory.
   As concerns the functional (4.3) one can show that under the same
conditions on the gauge group (4.7) and on the gauge field (4.8) in
even D case, the (4.3) is the total derivative in view of closure of
the form
\begin{eqnarray}
&&d\omega_{(2)} = 0\nonumber\\
&&\omega_{(2)} = Tr (F \wedge F)
\end{eqnarray}
  Hence integral over the form $\omega_{(2)}$ (4.3) is not  an action but is
an analog of the Pontryagin class. It is well known, that given the
closed invariant differential form
\begin{eqnarray}
&&d\omega=0, \nonumber\\
&&\Delta_{\alpha} \omega = 0
\end{eqnarray}
one can determine the Chern-Simons term
\begin{equation}
\omega = d \Gamma_{cs}
\end{equation}
For the form $\omega_{(2)}$ (4.11) the Chern-Simons term $\Gamma_{cs}$ has
the form
\begin{equation}
\Gamma_{cs} = AdA - {2 \over 3} A ( \eta A) A
\end{equation}
independently of the value of D. Notice that both $\omega_{(2)}$ and
$\Gamma_{cs}$
have sense in arbitrary dimension.

   Now one can write the gauge invariant Chern-Simons action for every
closed (D-1)-dimensional manifold $M^{D-1}$ :
\begin{equation}
S_{cs} = \int_{M^{D-1}} \Gamma_{cs}
\end{equation}
This action exists in arbitrary space-time dimension starting with
dimension 3. Such a generalization of the Chern-Simons term was
suggested in [10]. It is also particular result in our general
construction.

   To remind, for the standard Yang-Mills gauge fields (one-forms) the
Chern-Simons terms exist only in odd dimension (2k-1) and are
determined as follows:
\begin{equation}
d\check{\Gamma}_{cs} = \check{\omega}_{(2k)}
\end{equation}
where  $\check{\omega_{(2k)}}$  is closed invariant 2k-form:
\begin{equation}
\check{\omega}_{(2k)} = Tr [ F^{k}]
\end{equation}
here F is the curvature 2-form.

   The form $\omega_{(2)}$ (4.11) is similar to $\check{\omega}_{(4)}$. However
$\omega_{(2)}$ comprise components of any rank $N\geq 4$ , so the generalized
Chern-Simons term $\Gamma_{cs}$ (4.14) exists in any dimension $D\geq 3$ and,
in particular, in even dimension D, contrary to the standard gauge fields case
(the four-dimensional Chern-Simons example will be considered in the next
Section).

  However, the above discussion doesn't exhaust the rich structure of
NGATF Chern-Simons terms. Indeed, in dimension  $D\geq 6$ one can consider
the D-form:
\begin{equation}
\omega_{(3)} = Tr( F\wedge F \wedge F)
\end{equation}
which is closed and invariant under the same conditions (4.7). Hence
in high dimension D along with (4.14) there exists also the following
Chern-Simons term $\Gamma_{cs}^{(3)}$ :
\begin{equation}
d \Gamma^{(3)}_{cs} = \omega_{(3)}
\end{equation}
Further, in higher dimension D any invariant closed D-form:
\begin{equation}
\omega_{(k)} = Tr F^{k}
\end{equation}
determines the corresponding Chern-Simons   $\Gamma^{(k)}_{cs}$ :
\begin{equation}
d \Gamma^{(k)}_{cs} = \omega_{(k)}
\end{equation}
Any such term $\Gamma^{(k)}_{cs}$ defines the gauge invariant NGATF action
\begin{equation}
S^{(k)} = \int_{M^{D-1}} \Gamma^{(k)}_{cs}
\end{equation}
For a closed (D-1)-dimensional manifold $M^{D-1}$ .

   As a direct generalization one can also consider the NGATF taking
values in an Grassman algebra $\Lambda$ and write the corresponding
generalized Chern-Simons terms $\Gamma^{(k)}_{cs}$. In three dimension
we then obtain the field theory whose partition function is the Casson
invariant [11,12]. Hence, one can also expect the appearance of relevant
invariants under quantization of these super Chern-Simons actions in
arbitrary dimension.

   Summarizing, we obtained very rich Chern-Simons terms structure for
NGATF. One can use these along with (4.2) as actions in order to
determine the dynamical description of NGATF.

\section{The High-Dimensional Chern-Simons}
\setcounter{equation}0
   In three dimensions (4.14) gives us usual Chern-Simons term for
Yang-Mills field [5].
   As an next example let us consider the generalized Chern-Simons term in
four dimensions. For standard Yang-Mills gauge fields such a term
doesn't exist in D=4.

   From our general expression (4.14) we obtain:
\begin{equation}
\Gamma_{cs} = Tr(A_{(1)} \wedge dA_{(2)} + A_{(2)} \wedge dA_{(1)}
 + 2A_{(1)} \wedge A_{(1)} \wedge A_{(2)})
\end{equation}
It is easy to see that integral of (5.1) over a four-dimensional
closed manifold $M^{4}$
\begin{equation}
S_{cs} = \int_{M^{4}} \Gamma_{cs}
\end{equation}
coincides with
\begin{equation}
S_{cs} = \int_{M^{4}} 2Tr[ F_{(2)} \wedge A_{(2)}]
\end{equation}

and it is thus invariant under the gauge transformations (2.9):
\begin{eqnarray}
&&\delta_{\alpha} A_{(1)} = -d \alpha_{(0)} + \alpha_{(0)} A_{(1)}
- A_{(1)} \alpha_{(0)},\nonumber\\
&&\delta_{\alpha} A_{(2)} = -d \alpha_{(1)} - \alpha_{(1)} \wedge A_{(1)}
-A_{(1)} \wedge \alpha_{(1)} + \alpha_{(0)} A_{(2)} - A_{(2)} \alpha_{(0)}
\label{eq.6}
\end{eqnarray}
The Chern-Simons (5.3) with respect to (5.4) is transformed by the
total differential.

   It is worth noting that the action (5.3) is already well known in
the literature as the BF-system. This model was proposed in [6] (see
also [11,12]) as an example of an exactly soluble diffeomorphism
invariant theory. (A similar action was also considered by Freedman
and Townsend in [8]). As shown in [6], this model when quantized
describes the topological invariants such as linking numbers. The
transformations (5.4) contrary to [6] are not accidental but appear
as a part of more general and rich structure.

   As another example, let us consider the gauge gravity. The
corresponding generalization of the Lorentz group will be described by
(see Appendix for more details):
\begin{equation}
\xi G^{k}_{\ i} \wedge \eta_{kl} G^{l}_{\ j} = \eta_{ij}
\end{equation}
where $G^{i}_{\ j} = \sum_{p=0}^{D} G^{i}_{(p)\ j}$ is a matrix-valued
inhomogeneous differential form; $\eta_{ij}=diag(+1,-1,...,-1)$ is
the Minkowski D-dimensional metrical tensor,  $i,j =1,2,...,D$.

   Localizing the group (5.5) along the lines discussed in Sect.2,3,
one introduces the generalized k-forms connections $\omega_{(k)\ j}^{i}$.
However, not all of these are dynamically present in the Chern-Simons
action. In four dimensions, for example, one has the expression
similar to (5.3):
\begin{eqnarray}
&&S_{cs} = \int_{M^{4}} \Gamma_{cs},\nonumber\\
&&\Gamma_{cs} = Tr[ d \omega_{(1)} \wedge \omega_{(2)} + \omega_{(1)}
\wedge \omega_{(1)} \wedge \omega_{(2)}]
\end{eqnarray}
This action was also considered in details by Horowitz [6].

   It is worth noting that the group (5.5) is a super-generalization
of the Lorentz group. Hence the corresponding gauge gravity theory may
be treated as the new version of a supergravity theory.

   Now let us write precisely the generalized Chern-Simons term (4.14)
for closed manifold $M^{D}, D \geq 5$ :
\begin{center}
{\large\bf D=5}
\end{center}
\begin{equation}
S_{cs} = \int Tr[ A_{(3)} \wedge F_{(2)}]
\end{equation}
\begin{center}
{\large\bf D=6}
\end{center}
\begin{eqnarray}
&&S_{cs} = \int Tr[ F_{(2)} \wedge A_{(4)} - F_{(3)} \wedge A_{(3)} -
{1 \over 3} A_{(2)} \wedge A_{(2)} \wedge A_{(2)} ],\nonumber\\
&&F_{(3)}= d A_{(2)} + A_{(1)} \wedge A_{(2)} - A_{(2)} \wedge A_{(1)}
\end{eqnarray}
\begin{center}
{\large\bf D=7}
\end{center}
\begin{eqnarray}
&&S_{cs} = \int Tr[ A_{(5)} \wedge F_{(2)} + A_{(3)} \wedge
F_{(4)}],\nonumber\\
&&F_{(4)} = dA_{(3)} + A_{(3)} \wedge A_{(1)} + A_{(1)} \wedge A_{(3)}
\end{eqnarray}
\begin{center}
{\large\bf D=8}
\end{center}
\begin{eqnarray}
&&S_{cs}= \int Tr[ A_{(6)} \wedge F_{(2)} + A_{(5)} \wedge F_{(3)} + A_{(4)}
\wedge F_{(4)} + A_{(2)} \wedge A_{(3)} \wedge A_{(3)},\nonumber\\
&&F_{(3)} = dA_{(2)} + A_{(1)} \wedge A_{(2)} - A_{(2)} \wedge
A_{(1)}\nonumber\\
&&F_{(4)}= d A_{(3)} + A_{(1)} \wedge A_{(3)} + A_{(3)} \wedge A_{(1)} -
A_{(2)} \wedge A_{(2)}
\end{eqnarray}
   One can see that the generalized Chern-Simons actions in D=6,8 don't
have form of the B-F system in these dimensions. It should be noted that
though our Chern-Simons action for closed manifold $M^{D}, D=4,5,7,...$
has  the form of B-F system, it differs on B-F system for manifold with
boundary which is appropriate for canonical quantization of the model.
For open manifold this action is similar to standard Chern-Simons term
for Yang-Mills field in three dimensions [5].

\section{ Further generalizations and development}
\setcounter{equation}0
   There is a wide field for further generalizations and applications
of the suggested theory. Here we would like to mention some, in our
opinion, most interesting.

6.1. {\large \bf Generalization of gauge group}

   The gauge group in Sect.3 (see (3.9)) is the super-generalization
of the unitary group U(N). In general case, one can take any group g
(orthogonal, simplictic and etc.), which satisfies the condition
\begin{equation}
g^{\star} \eta g = \eta
\end{equation}
where $\star$ and $\eta$  are appropriate conjugation and invariant metric
tensor.
   Then one can determine the  group on differential forms
as follows:
\begin{equation}
\xi G^{\star} \wedge \eta G = \eta
\end{equation}
which will give us the superization of group $g$.

   Moreover there  is a further generalization  of (6.2) [13]. Indeed,
let $P$ be  an inhomogeneous differential form with  values in an matrix
space.  Let us  consider the  differential form  G, which satisfy the
condition:
\begin{equation}
\xi G^{\star} \wedge P \wedge G = P
\end{equation}
The differential form $P$ plays the role of fundamental form on a
manifold M which defines the scalar product for matter fields. In
general case there are two natural fundamental forms: zero-form and
D-form of volume. If $P$ is a zero-form  $P_{ij}$  one obtains the
definition (6.2). On the other hand if $P$ is a volume form then G in
(6.3) has only zero rank component $G_{(0)}$. Hence (6.3) is equivalent
to eq.(6.1) in this case.

   Notice that in these two described cases (a zero-form $P$ and a
volume-form $P$) the form $P$ belongs to the cohomology space $H^{(0)}(M)$
and $H^{(D)}(M)$ respectively. In general case one may assume that P is
closed form, $dP = 0$, and represents a relevant cohomology classes
\begin{eqnarray}
&&P_{ij} = \sum_{k=0}^{D} P^{(k)}_{ij} \nonumber\\
&&P_{ij}^{(k)} \in H^{(k)}(M)
\end{eqnarray}
For example, if M is a compact Kahler manifold then the even
cohomology groups  $H^{(2k)}(M)$ are not trivial [14]. So the
construction (6.3), (6.4) has a nontrivial realization.

   This superization of a group g (6.1) seems to be interesting since
it inherits the properties of initial classical group g (6.1) and
feels the geometry of underlying manifold $M^{D}$ .

6.2.{\large\bf Quantization and new topological invariants}

   It is well known that quantization of diffeomorphism invariant
metric independent actions reproduces topological invariants in terms
of quantum expectation values [6,7]. One obtains the Ray-Singer
torsion when quantizing AGATF [7] and linking numbers for BF-system
described in Sect.5 [6]. All these theories are particular cases of
the above considered NGATF theory. There exist two kinds of actions
for NGATF:
\begin{equation}
(F,F)= \int_{M^{D}} Tr (\tilde{F} \wedge F)
\end{equation}
and a wide class of the Chern-Simons actions
\begin{equation}
S^{(k)}_{cs} = \int_{M^{D}} \Gamma^{(k)}_{cs}
\end{equation}
Both of these actions are metric independent, so one can expect that
the topological invariants arise as a path integrals with any of these
actions:
\begin{equation}
Z({\cal O}_{1}, ..., {\cal O}_{j}) = \int {\cal D} A {\cal O}_{1}...
{\cal O}_{j} \exp (-S[A])
\end{equation}
where ${\cal O}_{1}, ..., {\cal O}_{j}$ are gauge invariant operators.
In the abelian case one usually takes as ${\cal O}_{i}$ integrals over
the k-dimensional cycles $\gamma_{(k)}$ [6]:
\begin{equation}
{\cal O} = \int_{\gamma_{(k)}} A_{(k)}
\end{equation}
For the one-form $A_{(1)}$ one has the gauge invariant operator [5]
\begin{equation}
{\cal O}= Tr Pexp \int_{\gamma_{(1)}} A_{(1)}
\end{equation}
which is holonomy of connection $A_{(1)}$ around loop $\gamma_{(1)}$.
In our paper [15] we find the corresponding generalization of these formulas
for NGATF.

    Notice that actions (3.21) for matter fields are also metric
independent. Hence they are also subject for quantization leading to
topological invariants.

6.3 {\large\bf Physical models with NGATF}

   The abelian gauge antisymmetric tensor fields play an important
role in the supergravity, in the gauge theory of gravity and in the
string theory [1]. The NGATF theory is generalization of both AGATF
and the standard gauge fields. Hence NGATF may find a wide
applications.

   As it was shown in Section 3. the supergroup appears naturally in
our approach. In fact, it is not necessary to introduce an additional
grassman variables in the theory. The superspace has a natural
realization as a space of variables {$x^{\mu}, dx^{\nu}$}, i.e. the
space $\Lambda^{\ast}(M)$. Hence this approach is appropriate for
description of supergauge theory. For instance, one may attempt in
construction of the super-Poincare gravity in the framework of gauge
approach with NGATF. In particular, it allows to write the gauge
gravity action without a help of metric using only the gauge fields.
Such a theory would describe a phase in which general covariance is
unbroken. The metric will appear as a Goldstone boson of a spontaneously
broken (local) general covariance, as discussed in [16].

   It should be noted here that the possibility to describe gravity
with help of SL(3)-algebra valued two-form instead of metric was also
considered in [17].

   Another promising application of NGATF is the string field theory.
There an approach was developed which gives a nonperturbative
description of string with the help of infinit-dimensional analog of
the Chern-Simons action [18]. One may consider the corresponding gauge
field as the inhomogeneous differential form on the space of string
configurations. The appropriate gauge group would be the
diffeomorphism group Diff($S^{1}$) probably together with the Kac-Moody
extension of the Poincare group [19].

   One may also expect the NGATF to play a role in condensed matter
systems where the standard Chern-Simons approach has proved to be very
useful [20].

{\large\bf Acknowledgements}\\
   I am grateful to Yu.N.Obukhov for valuable discussions and careful
reading the manuscript. I also thank D.Leites, V.Serganova, D.Fursaev
and O.Zakharov for conversations.

{\appendix \noindent{\large \bf Appendix.} \\
\def\theequation{A.\arabic{equation}}
\setcounter{equation}{0}
   Let us obtain the general solution for the equation
\begin{equation}
\xi G^{+} \wedge G = 1
\end{equation}
which determines the gauge group element G:
\begin{equation}
G^{i}_{\ j} = \sum_{k} G^{i}_{(k) j}
\end{equation}
In general case G can be represented in the form
\begin{equation}
G= g_{0} h ,
\end{equation}
where $g_{0}$    is a matrix-valued zero-form, and $h$ is a matrix valued
inhomogeneous differential form with a zero rank component equal to
$h_{(0)}  =1$.

  Inserting (A.3) into (A.1) we assume that the matrix $g_{0}$ is unitary:
\begin{equation}
g^{+}_{0} g_{0} = 1
\end{equation}
Then we obtain the equation for $h$:
\begin{equation}
\xi h^{+} \wedge h = 1,\\\ h_{(0)} =1
\end{equation}
Any form $h$ with $h_{(0)}  =1$ is decomposed into a sum
\begin{equation}
h = 1 + H + {\cal F}
\end{equation}
where the forms H and ${\cal F}$   are both of rank $\geq 1$ and satisfy the
conditions:
\begin{equation}
\xi H^{+} = - H, \\\ \xi{\cal F}^{+} = {\cal F}
\end{equation}
{}From (A.6) and (A.7) we obtain that
\begin{equation}
h^{-1} = \xi h^{+} = 1 - H + {\cal F}
\end{equation}
Inserting this expression into (A.5) one finds
\begin{equation}
{\cal F} \wedge {\cal F} + 2 {\cal F} - H \wedge H = H \wedge {\cal F} -
{\cal F} \wedge H
\end{equation}
It is a consequence of eqs.(A.5)-(A.8) that
\begin{eqnarray}
&&H={1 \over 2} ( g - g^{-1}), \nonumber\\
&&{\cal F}= {1 \over 2} (g + g^{-1}) - 1
\end{eqnarray}
One concludes from this that H and ${\cal F}$   commute:
\begin{equation}
H \wedge {\cal F} = { \cal F} \wedge H
\end{equation}
The last equation and conditions (A.7) suggest representing ${\cal F}$   as
an even polynomial of H:
\begin{equation}
{\cal F} = \sum_{k} a_{k} H^{2k}
\end{equation}
Substituting (A.12) into (A.9) and taking into account (A.11) one
gets:
\begin{equation}
\sum_{k} 2a_{k} H^{2k} + \sum_{p,m} a_{p} a_{m} H^{2(p+m)} - H^{2} = 0
\end{equation}
Then one obtains the following recurrent identities for $a_{k}$ :
\begin{eqnarray}
&&a_{1} = {1 \over 2} \nonumber\\
&&a_{2} = - {1 \over 8} \nonumber\\
&&a_{k} = - {1 \over 2} \sum_{p=1}^{k-1} a_{p} a_{k-p},  k \geq 2
\end{eqnarray}
   Then the parameter space of the group $G$ (A.1) consists of
parameters of unitary matrix $g_{0}$ (A.4) and independently of
differential form
\begin{equation}
H= \sum_{k\geq 1}{ (\imath)^{k(k-1) \over 2} \over k!}
H^{a}_{\mu_{1}...\mu_{k}}
\lambda ^{a} dx^{\mu_{1}}\wedge ... \wedge dx^{\mu_{k}},
\end{equation}
where $\{\lambda^{a}\}$ is a set of antihermitean matrices.

   It is easy to obtain the group multiplication law for H. We should
note here that this is nonabelian even in the abelian case (when
$g_{0} \in U(1)$).

   Let us consider in brief the case of the Lorentz group. The gauge
group is determined in this case by the condition:
\begin{equation}
\xi G^{T} \wedge \eta G = \eta
\end{equation}
where $\eta_{ij} = diag (+1,...,-1)$ is the metric tensor, and $ G = \sum
G^{i}_{(k) j}$ is matrix-valued differential form. The element G (A.16)
can also be represented in the form
\begin{equation}
G = g_{0} ( 1 + H + { \cal F }),
\end{equation}
where $g_{0}$ is the Lorentz group element ($g_{0}^{T}  \eta g_{0} = \eta$),
and H and ${ \cal F }$ satisfy
\begin{eqnarray}
&&\xi H_{ij} = - H_{ji}\nonumber\\
&&\xi { \cal F}_{ij} = {\cal F }_{ji}
\end{eqnarray}
where we denote  $ H_{ij} = H^{k}_{\ j} \eta_{ik}$.

   As earlier the form $ {\cal F}$ can be represented as polynomial of H
(A.12).
{}From (A.18) one obtains for the k-rank component $H_{(k)}$:
\begin{equation}
H^{(k)}_{ij}= - (-1)^{ k(k-1) \over 2} H^{(k)}_{ji}
\end{equation}
In four-dimensional case we have
\begin{eqnarray}
&&H^{(k)}_{ij} = - H^{(k)}_{ji}, k=0,1,4\nonumber\\
&&H^{(k)}_{ij} =\ H^{(k)}_{ji}, k=2,3
\end{eqnarray}
Hence H is not decomposed with respect to basis of the Lorentz group
generators, as we had in unitary group case (see eq.(A.15)). One
should consider additionally the symmetric matrices basis for
$H^{(k)}, k=2,3$. This is the main difference in the definition of the
group G for complex and real cases.
}

\end{document}